\documentclass[twocolumn]{aa}			
\usepackage{graphicx}

\usepackage{txfonts}
\usepackage{natbib}
\bibpunct{(}{)}{;}{a}{}{,}
\def\gapprox{{_>\atop{^\sim}}}
\def\lapprox{{_<\atop{^\sim}}}

\def\cmmd{\rm {cm^{-3}}}

\def\s-1{\rm {s^{-1}}}
\def\twco{$^{12}$CO}
\def\thco{$^{13}$CO}

\def\kms {\hbox{${\rm km\,s}^{-1}$}}
\def\msun{M$_{\odot}$}

\def \r{${\cal R}$}
\def\farcs{\hbox{$.\!\!^{\prime\prime}$}}

\begin{document}
   \title{\thco\ 1--0 imaging of the Medusa merger, NGC~4194}

   \subtitle{Large scale variations in molecular cloud properties}

   \author{S. Aalto\inst{1}
        \and
        R. Beswick\inst{2}
	\and
	E. J\"utte\inst{3}
          }

   \offprints{S. Aalto}

   \institute{Department of Earth and Space Sciences, Chalmers University of Technology, Onsala Observatory,
              SE-439 94 Onsala, Sweden\\
              \email{saalto@chalmers.se}
        \and
Jodrell Bank Centre for Astrophysics,
School of Physics and Astronomy, The University of Manchester
Oxford Road, Manchester M13 9PL, UK
	\and
	University of Bochum, Department of Astronomy, 44780 Bochum, Germany
             }

   \date{Received x; accepted y}


    \abstract
  {}
   {Studying molecular gas properties in merging galaxies gives important clues to 
the onset and evolution of interaction-triggered starbursts. The $\frac{^{12}{\rm CO}}{^{13}{\rm CO}}$ line intensity
ratio can be used as a tracer of how dynamics and star formation processes impact the gas properties.
The Medusa (NGC~4194) is particularly interesting to study since its ${L_{\rm FIR}\over L_{\rm CO}}$ ratio rivals that of ultraluminous galaxies (ULIRGs), despite the comparatively modest luminosity, indicating an exceptionally high star formation 
efficiency (SFE) in the Medusa merger.}
   {High resolution OVRO 
(Owens Valley Radio Observatory) observations of the $^{13}$CO 1--0 have 
been obtained and compared with matched resolution OVRO $^{12}$CO 1--0 data 
to investigate the molecular gas cloud properties in the Medusa merger.
}
   {Interferometric observations of \twco\ and \thco\ 1--0 in the Medusa (NGC~4194)
merger show the ${{^{12}{\rm CO}} \over {^{13}{\rm CO}}}$ 1--0 intensity ratio (\r) increases
from normal, quiescent values (7-10) in the outer parts ($r>$ 2 kpc) of the galaxy to high (16 to $>$40) values in the
central ($r<$ 1 kpc) starburst region.  In the central two kpc there is an east-west gradient in \r\ where
the line ratio changes by more than a factor of three over 5\arcsec\ (945\,pc). The integrated
\thco\ emission peaks in the north-western starburst region while the central \twco\ emission is strongly
associated with the prominent crossing dust-lane. 
}
   {We discuss the central east-west gradient in \r\ in the context of gas properties in the starburst and
the central dust lane. We suggest that the central gradient in \r\ is mainly caused by diffuse gas in
the dust lane. In this scenario, the actual molecular mass distribution is better traced by the \thco\ 1--0
emission than the \twco. The possibilities of
temperature and abundance gradients are also discussed. We compare the central
gas properties of the Medusa to those of other minor mergers and
suggest that the extreme and transient phase of the Medusa star formation activity has
similar traits to those of high-redshift galaxies. 
}
\titlerunning{Molecular clouds in NGC~4194}
\maketitle

%

\section{Introduction}

Studying the molecular cloud properties in starburst, active and interacting
galaxies is important in order to understand the feedback mechanisms between
star formation, dynamics and the interstellar medium. Important probes of cloud
properties include molecules such as HCN, HCO$^+$ and HNC
\citep[e.g.][]{krips08,gracia08,imanishi04,gao04,aalto02} that trace the dense ($n \gapprox 10^4$ $\cmmd$)
star forming phase of the molecular gas. 

Alternatively, to study the bulk properties of the molecular gas one can use the
ratio between \twco\ and its isotopomer \thco. Globally, there is a correlation between
the \twco/\thco\ 1--0 ratio (\r) and the FIR $f$(60$\mu$m)/$f$(100$\mu$m) flux ratio 
\citep[e.g.][]{young86, aalto95}. The extreme values of \r\ (i.e. \r$>$20)
generally occur in luminous merging galaxies with large dust temperatures \citep[e.g.][]{aalto95,glenn01}.
Within galaxies there is a general trend of increasing \r\
towards the central region where the gas is warmer and denser \citep[e.g.][]{wall93,
aalto95, paglione01}. 

Both high kinetic temperatures and large turbulent
line widths will decrease the optical depth ($\tau$), of the \twco\ and \thco\ 1--0 lines and thus elevate the line ratio \r.
Therefore, information on the spatial variation of \r\ can be used to identify regions
of extreme or unusual physical conditions in the molecular gas - and global values can help classify galaxies.
\citet{aalto95} suggested some general diagnostics of the cloud conditions and environment based on global values of \r: 
\begin{itemize}

\item Small ratios, \r \, $\approx 6$ are an indication
of a normal Galactic disc population of clouds dominated by cool, self-gravitating Giant Molecular Clouds (GMCs). 

\item Intermediate ratios
$10 \lapprox {\cal R} \lapprox 15$ are associated with the inner regions of normal starburst
galaxies where the gas is warmer and denser than in the disc.

\item  The extreme values \r\ $> 20$ seem to originate in warm, turbulent, high pressure gas in 
the centres of luminous mergers with highly compact molecular distributions and gas surface densities
in excess of 10$^{4}$~\msun\ pc$^{-2}$---two orders of magnitude higher than in typical Milky Way 
GMCs. Large surface densities require high
pressures in hydrostatic equilibrium and low density ($n < 10^3$ $\cmmd$) - \twco\ emitting - gas
must be supported by large turbulent line widths ($P \propto n (\delta V)^2$). Thus, $\tau_{\rm CO}$ can be reduced to moderate
($\approx 1$) values, resulting in large \r. 

\end{itemize}

\noindent
More recent studies show that high values of \r\ also occur more locally in
less extreme galaxies. \citet{tosaki02} find this in the spiral arms of M~51 and \citet{meier04} in the centre 
of the spiral galaxy NGC~6946. \citet{huette00} find large values of \r\ in the large
scale bar of NGC~7479 as a result of the dynamical impact of a density wave on gas properties.\\

The nearby merger NGC~4194 - the Medusa merger - belongs to a class of lower luminosity 
($L_{\rm FIR} = 8.5 \times 10^{10}$ L$_{\odot}$ at $D$=39 Mpc)
E+S mergers \citep{aalto00, manthey08}, an order of magnitude fainter than well known Ultra Luminous Galaxies (ULIRGs
($L_{\rm FIR} \gapprox 10^{12}$ L$_{\odot}$))
such as Arp~220. The Medusa has an elevated value of \r\ ($\approx$20) \citep[e.g.][]{aalto91b, casoli92, glenn01},
but in contrast to the more luminous
high-\r\ mergers, NGC~4194 has a relatively extended (2~kpc) starburst region \citep[e.g.][]{armus90, prestwich94}
and \citet{aalto00} found that the molecular
gas was also distributed on the comparatively large scale of 25$''$ (4.7 kpc), despite its advanced stage of merger.
Due to the linear extent of the molecular gas and relative 
proximity of the Medusa merger, it is possible to spatially resolve the 
relative distributions of \twco\ and \thco\ and hence allow the investigation 
of the underlying causes of the elevated \r\ values observed in this galaxy.
The Medusa starburst is particularly interesting to study since its ${L_{\rm FIR}\over L_{\rm CO}}$ ratio of 163 rivals
that of ULIRGs (for example Arp~220 with a ${L_{\rm FIR}\over L_{\rm CO}}$ of 210), despite the comparatively
modest luminosity, indicating an exceptionally high star formation efficiency (SFE) in the Medusa merger.
\citet{gao04} find that for galaxies with ${L_{\rm FIR} \lapprox 10^{11}}$ L$_{\odot}$
the ${L_{\rm FIR} \over L_{\rm CO}}$ ratio is typically 33. At higher luminosity they find that ${L_{\rm FIR} \over L_{\rm CO}}$
instead increases to 100--300. \citet{gao04} suggest that, in contrast to CO, the ${L_{\rm FIR} \over L_{\rm HCN}}$ ratio 
is independent of luminosity with ${L_{\rm FIR} \over L_{\rm HCN}}$=900. Interestingly the Medusa deviates significantly
from this HCN correlation with a ${L_{\rm FIR }\over L_{\rm HCN}} \gapprox 4075$ based on upper limits from \citet{aalto00}).
Further discussion of the underlying causes of this discrepancy are presented in \citet{aalto00}.

We have imaged NGC~4194 in the \thco\ 1--0 line with the Owens Valley Radio Observatory (OVRO) millimetre array.
Our aim was to compare the distribution of the \twco\ and \thco\ emission within the galaxy and to see how
it relates to star formation, gas surface density and dynamics. 
In sections 2 and 3 we discuss the observations and results. In section 4 we discuss the \r\
line ratio variations within the Medusa merger in terms of warm gas in the starburst region and
possible diffuse gas in the central dust lane. Future observational tests for these 
proposed scenarios are suggested.


\section{Observations}

We have obtained maps of \thco\ 1--0 using
the Caltech six-element OVRO millimeter array. Two tracks were taken in the low resolution configuration
in April 2000. The naturally weighted synthesised beam size is 
$4\hbox{$\,.\!\!^{\prime\prime}$}56 \times 3\hbox{$\,.\!\!^{\prime\prime}$}98$
($472 \times 378$ pc for $D$=39 Mpc) and the beam position angle (BPA) is --40$^{\circ}$.

System temperatures were 500--600 K. The quasar 1150+497 was 
used for phase calibration and Neptune and Uranus for
absolute flux calibration.  
The digital correlator, centred at $V_{\rm LSR} = 2560$ \kms, provided
a total velocity coverage of 1170\,\kms.
Data were binned to 4 MHz resolution, corresponding to 
11 \kms, and to construct the map, the resolution was reduced to 44 \kms.
The sensitivity of this map is 3\,mJy\,beam$^{-1}$ channel$^{-1}$,
corresponding to 0.014\,K\,channel$^{-1}$. 

\begin{figure}
\label{f:cont}
\resizebox{5cm}{!}{\includegraphics[angle=0]{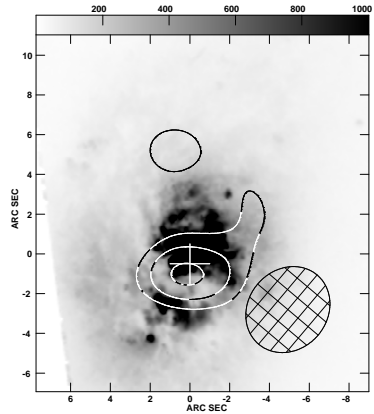}}
\caption{\label{f:cont} The 110\,GHz continuum of NGC~4194 overlaid on an {\it HST} WFPC2 image. The contours
are: $0.9 \times (-3,3,4,5)$ mJy\,beam$^{-1}$. Rms in the map is 1\,mJy\,beam$^{-1}$ thus the first contour is
at 2.7$\sigma$. The peak flux density is 4.6\,mJy\,beam$^{-1}$. The cross marks the position of
the 1.4\,GHz continuum \citep{condon90}. Hatched ellipse indicate beam size and orientation.}
\end{figure}

\subsection{Continuum subtraction}
\label{s:cont}

Using the {\sc aips} task {\sc imlin} we fitted continuum to the line-free
channels and subtracted from the cleaned map. 
The resulting continuum map is presented in Fig.~\ref{f:cont} 
and the flux density and position in Tab.~\ref{t:flux}. The 1.4 GHz continuum
position of \citet{condon90} agree within $0.''4$ with the 110 GHz position
which is within the positional errors of the \thco\ map.
The 1.4\,GHz radio continuum emission is dominated by synchrotron emission
resulting primarily from star-formation, along with a small contribution
from the compact core \citep{beswick05,condon90}. The 110\,GHz the
emission (see Fig.~\ref{f:cont}) is coincident with the area of highest dust
obscuration observed in the optical. This emission is tracing the thermal
emission from dust in this region illuminated by ongoing star-formation.
This is consistent with the alignment in the 110 and 1.4\,GHz continuum
positions observed.

\subsection{Alignment of the data}

Both the \twco\ and \thco\ observations were made using the same instrument 
and employing identical gain and passband calibrators, thus minimising 
potential misalignments due to instrumental effects. 
Furthermore, the 110~GHz continuum extracted from the \thco\
data set is positionally coincident with the 1.4~GHz continuum
(see Sect.~\ref{s:cont}) suggesting that there are no systematic
errors in the positions of the \thco\ data set.
We also note that the shift between \twco\ and \thco\ is partially
caused by missing \thco\ emission in some velocity channels (see Sect.~\ref{s:chan}).

\begin{table}
\caption{\label{t:flux} Continuum and \thco\ line results.}
\begin{tabular}{ll}
 & \\
\hline
\hline \\ 
Continuum$^a$: & \\
\, \, position (J2000) & $\alpha$:  12:14:09.66 \\
                           & $\delta$: 54:31:35.0 \\
\\
\, \, flux density(mJy) & $4.6 \pm 1$  \\

\\
Line: & \\
\, \, peak flux density & \\
\, \, (mJy\,beam$^{-1}$)& 25 $\pm$ 3 \\ 
\\
\, \, integrated flux density$^b$ &  \\
\, \, (Jy \kms) & $3.3 \pm 0.6$ (centre)\\
\, \, (Jy \kms) & $1.6 \pm 0.5$ (north) \\
\\
\hline \\
\end{tabular} 

a) A Gaussian was fitted to the continuum image. 
The error is 1$\sigma$ rms.
The 1.4\,GHz radio continuum position 
reported by \citet{condon90} is: $\alpha$: 12:14:09.66 \, $\delta$: 54:31:35.5 (J2000).

b) A Gaussian was fitted to the integrated intensity map of \thco. The
peak integrated flux is $1.9 \pm 0.3$ (Jy \kms) in the centre. Thus, the central source is slightly
resolved. The error is 1$\sigma$ rms.

\end{table}

\section{Results: Morphology of the \thco\ emission}

\begin{figure}
\resizebox{9cm}{!}{\includegraphics[angle=0]{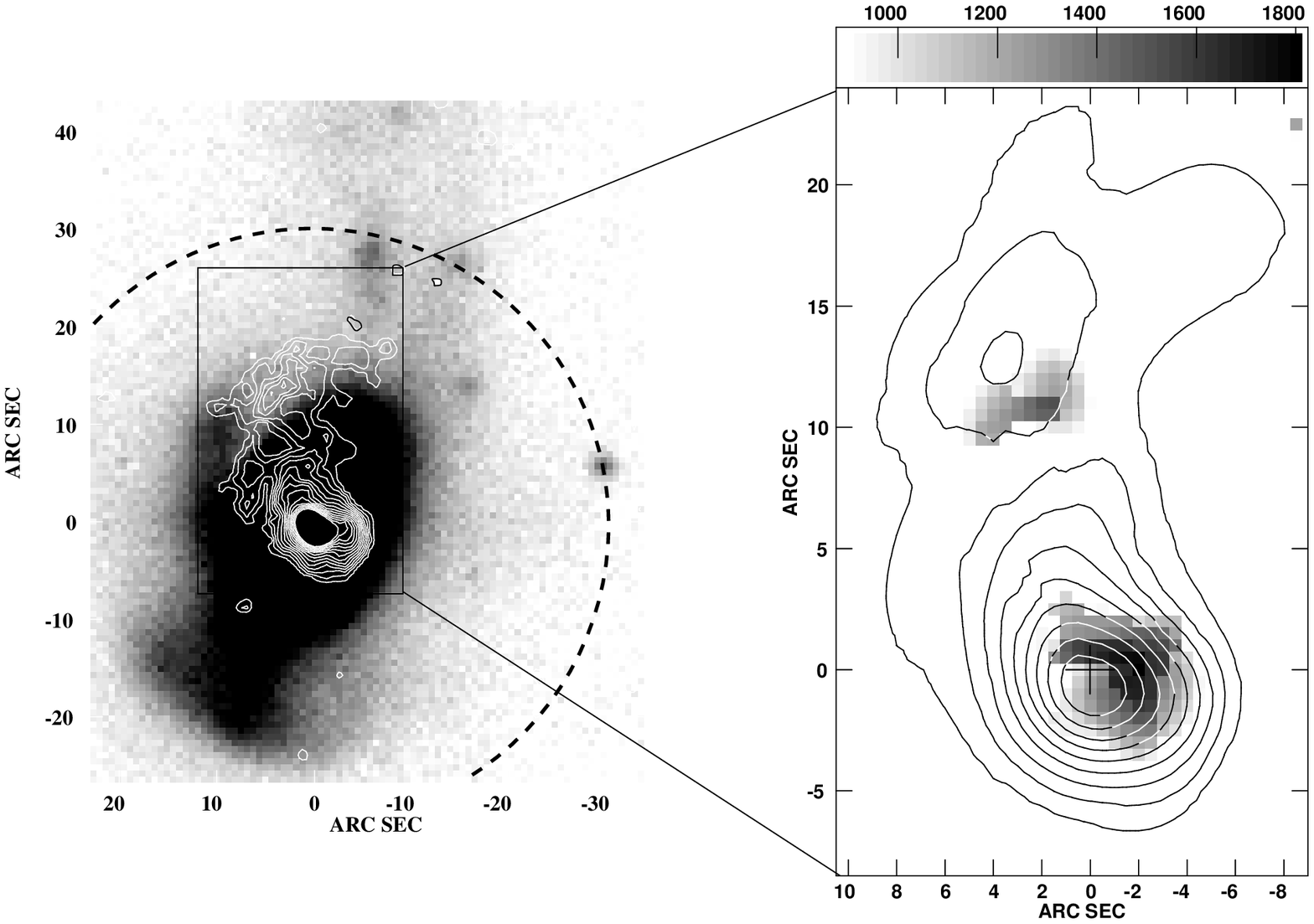}}
\caption{\label{f:co_opt} Left: Overlay of the \twco\ contours (white, apart from clouds in the
tail marked in black) on a greyscale, overexposed optical R-band image \citep{mazzarella93}.
The dashed curve marks the edge of the OVRO primary beam (from \citet{aalto00}). 
Right: Integrated \thco\ 1--0 line
emission in greyscale overlaid on a \twco\ contour map smoothed to the \thco\ resolution of $4.''5$.
The greyscale range is  0.9 -- 2.0 Jy \kms beam$^{-1}$. The CO contours are
5.3 $\times$ (1,2,3,4,5,6,7,8) Jy \kms beam$^{-1}$ (starting at 6$\sigma$).
Cross indicate position of 1.4 GHz radio continuum peak (as in Fig.~\ref{f:cont}).
}
\end{figure}

\begin{figure*}
\resizebox{\hsize}{!}{\includegraphics[angle=0]{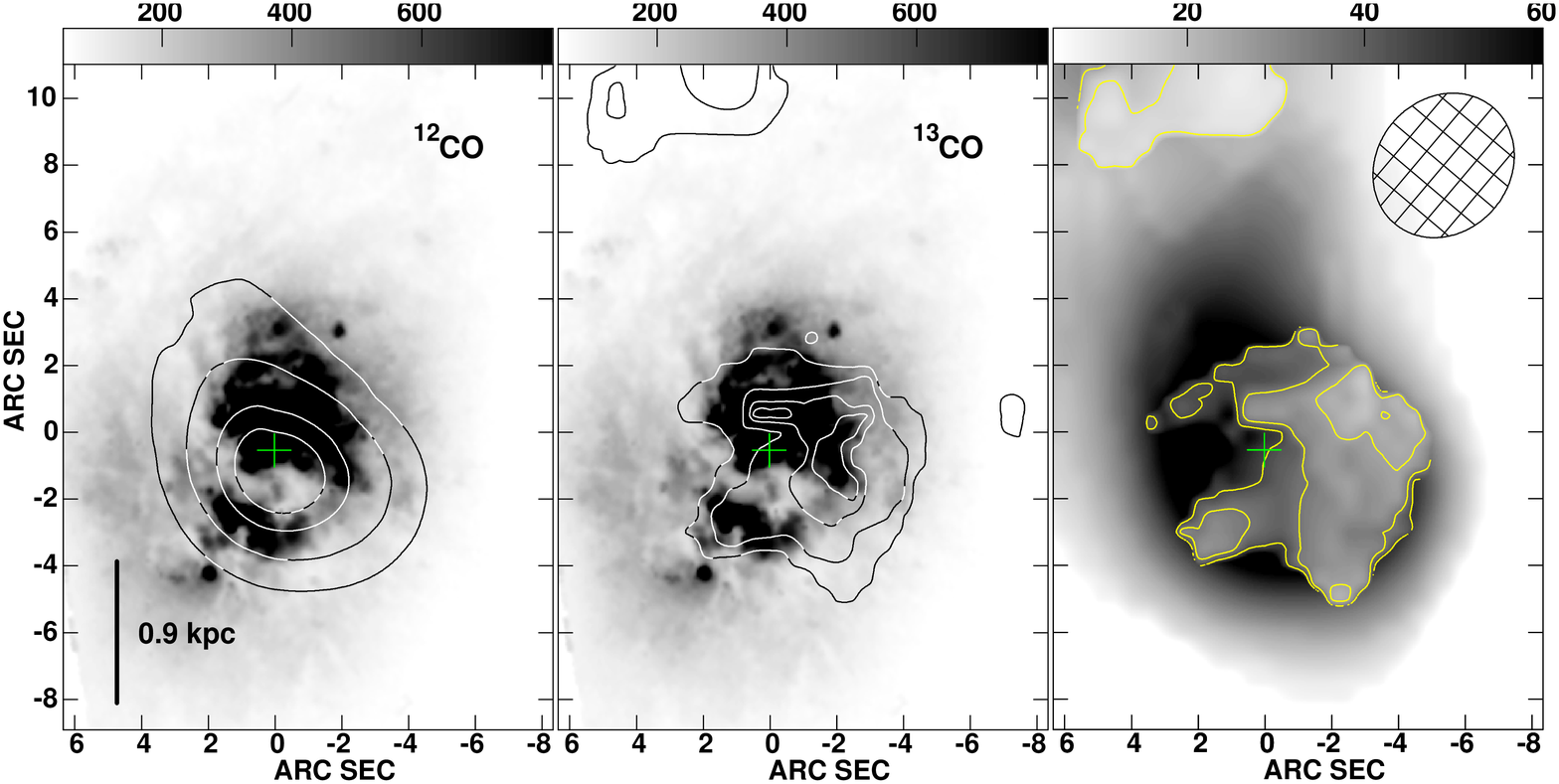}}
\caption{\label{f:over2} 
Left: \twco\ contours on a WFPC2 image zoomed in on the central region.
Centre: \thco\ contours on WFPC2 image. Both \twco\ and \thco\ contours are
40 \% of peak flux $\times$ (1, 1.5, 2, 2.25).  For \thco\ the first contour starts at
2.5$\sigma$, for \twco\ the first contour starts at 24 $\sigma$.
Right: Line ratio map where the greyscale ranges from light (\r=10) to dark (\r=60).
Both limits and real values in greys (same scale), and the area where there are actual
measures overlaid with contours (yellow). Contours are ratios \r = 15, 25, 35, 45. 
Limits are at the 3$\sigma$ \thco\ detection threshold.  Please note the small
pixel-size ($0.''5$) of the \twco\ and \thco\ maps which may give the impression that there
is structure beyond the actual angular resolution of the \twco\ and \thco\ maps. 
 }
\end{figure*}

\begin{figure*}
\resizebox{14cm}{!}{\includegraphics[angle=0]{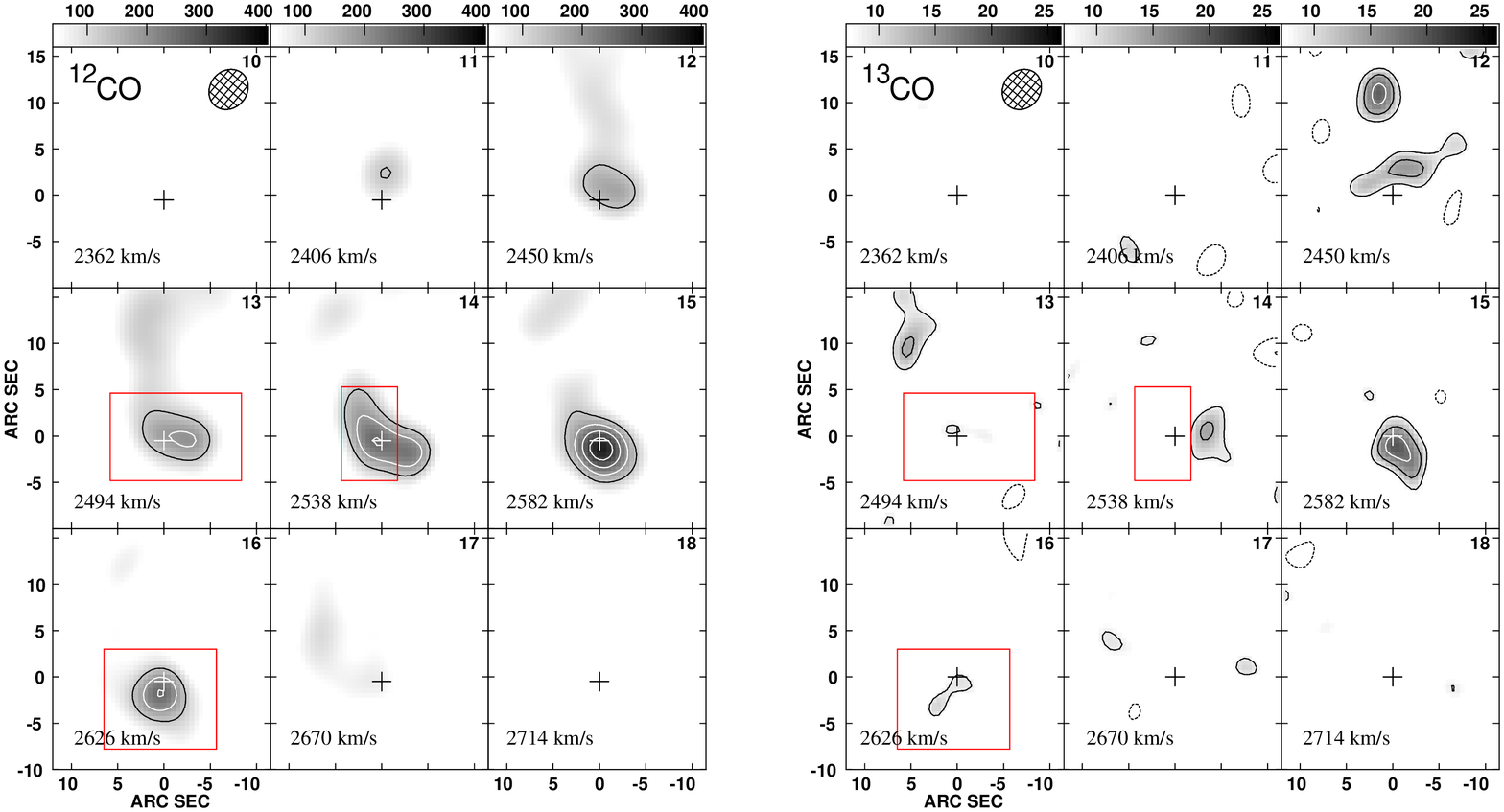}}
\caption{\label{f:chan} Channel maps of \twco\ (left) and \thco\ (right).
Boxes indicate channels of ``missing''  \thco\ emission compared to \twco. 
The \twco\ map has been smoothed to the same resolution as \thco.
Intensity levels for \twco\ and \thco\ are same fraction of peak: 35\% (-1.5,-1,1.5,2,2.5) -- 
where the peak is 360 and 25\,mJy\,beam${-1}$ for the \twco\ and
\thco\ respectively.
Note that for \twco\ a big fraction of the detected, extended flux is not shown due to the cut-off. The comparative channels maps show the position of the peak emission for \twco\ and \thco.
The velocity in panel 18=2714\,\kms, panel 14 is 2538\,\kms, and panel 10=2362\,\kms. Increment between panels is 44\,\kms. Crosses indicate the 1.4 GHz radio continuum position. The beam is marked in the upper right corner of the first panel for the
\twco\ and \thco\ channel maps. Greyscale for the \twco\ map ranges from 50 to 400\,mJy\,beam$^{-1}$ and for
\twco\ from 7.5 to 26\,mJy\,beam$^{-1}$. For the \thco\ map the first contour corresponds to 2.5$\sigma$. }
\end{figure*}

\subsection{Previous observations: The \twco\ distribution}

High resolution OVRO aperture synthesis maps of the
\twco\ 1--0 emission in the ``Medusa'' galaxy merger (NGC~4194) 
are presented in \citet{aalto00}. It was found that
the molecular emission is surprisingly extended. The 
\twco\ emission is distributed on a total scale of 25\arcsec\ (4.7\,kpc)
--- despite the apparent advanced stage of the merger.
The complex, striking \twco\ morphology 
occupies the centre and the north-eastern part of the main optical body
(see Fig.~\ref{f:co_opt} (left panel)).
The extended \twco\ flux density traces the two prominent dust lanes: 
one of which crosses the central region at right angles to the optical 
major axis, and the second which curves to the north-east and joins the 
base of the northern molecular tidal tail.
The total molecular mass in the system is estimated to be at most
$2 \times 10^9$ M$_{\odot}$,
depending on which \twco\ - H$_2$ conversion factor is applicable.

\subsection{Integrated \thco\ 1--0 intensity} 

The naturally weighted, $4.''5$ resolution, \thco\ emission is 
distributed on a scale of 8\arcsec\ (1.8 kpc) in an oval 
structure centred on the optically bright centre of the galaxy. There is also an off-centre \thco\
concentration  10\arcsec\ north of the centre. In Fig~\ref{f:co_opt} (right panel) we show
the integrated  \thco\ map overlaid on a \twco\ map smoothed to the \thco\
resolution. Integrated- and peak flux densities are listed in Tab.~\ref{t:flux}.

The \thco\ emission in the inner 8\arcsec\ is shifted towards the north-west
as compared to the \twco\ distribution. The shift is 2\farcs3 (430\,pc).
Also the northern \thco\ emission peak is shifted off the
corresponding \twco\ peak to the south-west by 2\farcs5. In Fig.~\ref{f:over2} (centre panel)
the \thco\ 1--0 map is overlaid on an {\it HST} WFPC2 image. This shows that
the \thco\ emission is not tracing the central dust lane (which the \twco\ does (left panel
in Fig.~\ref{f:over2}))
but instead is shifted towards the north-west part of the starburst and towards
the western dust features. The moment maps were produced through smoothing in velocity and space
(3 channels, 3 pixels boxcar smoothing) and then applying a 2$\sigma$ flux cut-off (allowing both
positive and negative flux levels). 
 
\subsection{The channel map}

\label{s:chan}

In Fig.~\ref{f:chan} the channel maps of \twco\ and \thco\ 1--0 are presented, smoothed
to a resolution of 5\arcsec. 
Compared to \twco, \thco-emission is ``missing'' primarily in three velocity bins in the centre: 2495, 2538 and 2620 \kms.
In the first case we find \thco\ emission 10\arcsec\ north of the centre - despite \twco\ in the same bin having
a peak in the centre. 
In the second case \thco\ emission is found only west of the centre while \twco\ is distributed east and west
of the radio peak. In the 2620 bin there is a bright \twco\ peak in the centre which is missing in \thco.
In general, the channel maps show that there is no \thco\ emission east, north-east of the radio continuum
peak apart from a small region of emission 10\arcsec\ north of the 
central region.

\subsection{Line ratios}

\label{s:ratio}

We find that \r\ is 7 -- 10 in the region 10\arcsec\
north of the centre - while in the centre it is significantly greater:
ranging from 16 in the western region of the centre to
$>$45 (3$\sigma$) in the eastern part of the centre
where the central dust lane is making a 90 degree turn. 
{\it Over a distance of 5\arcsec\ (945\,pc) the line ratio, \r, changes by more 
than a factor of 3.} A map of the varying \r\ line ratios is displayed in Fig~\ref{f:over2} (right panel).

\section{Discussion}

The line ratio change in the Medusa occurs on scales of 0.5 to 1 kpc
where the ratio goes from extreme ratios $>$40 to more normal starburst
ratios of 16. 
This observed gradient in \r\ is likely to be caused either by the 
presence of diffuse, unbound molecular gas or a temperature gradient in 
the ISM. In the following section we discuss each of these scenarios and 
compare these with observations in other galaxies. 

\subsection{Diffuse gas in the central dust lane?}

\label{s:diffuse}

Optical images of the Medusa merger are dominated by the strong absorption from several
dust lanes - the most prominent one crossing the central starburst region.
We suggest that the intense starburst activity is being fed by gas funnelled
along the dust lane - supported by the bright \twco\ emission associated
with the dust lane(s), and the distribution and dynamics seen in the H{\sc i} absorption 
at high resolution \citep{beswick05}.
The central dust lane is in front of the main burst region in the east and
then appears to curve into the burst region itself at the centre.
A possible scenario is then that the molecular gas is diffuse (non self-gravitating)
in the dust lane region - creating the general east-west gradient in the line ratio \r.

Interestingly, a comparison of the Medusa \thco\ image
and the H$\alpha$ images of \citet{hattori04}
show that the position of the H$\alpha$ equivalent width (EW) peak coincides 
with the regions of lower values of \r. The large EW is consistent with the region being
an extra-nuclear star forming region. It coincides with the western radio continuum
arm and the Giant Molecular Association (GMA) feature $c$ in the high-resolution 
\twco\ map (see \citet{aalto00}). Thus, it seems that the \thco\ emission correlates spatially
with more quiescent, spiral-arm like star formation, shifted from the dust lane and the 
central starburst. 
{\it Note that if the \twco\ 1--0 is mainly tracing diffuse gas the \thco\ 1--0 emission is
showing the real mass distribution of the molecular gas}.
Fig.~10 of \citet{aalto00} shows that the correlation between 1.4 GHz radio continuum and
the central dust lane is poor - this could be a further indication of both a lack of ongoing
star formation in the dust lane as well as the molecular gas potentially being diffuse in the 
dust lane (see for instance the discussion of \twco\ and radio continuum in the spiral arm
of M~83 by \citet{rand99}).

\subsubsection{Spatial shifts in \r\ in other galaxies}

Even if the central region of the Medusa is more chaotic
than that of a density wave spiral galaxy, and the extinction structure suggests
a three-dimensional central region, it is tempting to compare the shift in \twco\
and \thco\ to that found in other galaxies.  
\citet{tosaki02} found a spatial shift between \twco\ and \thco\ 1--0 emission on 
similar spatial scales in the southern spiral arm of M~51.
The shift seems to separate diffuse (unbound) lower density gas (as traced by \twco)
from self-gravitating gas emitting the bulk of the \thco\ line
emission. \citet{tosaki02} suggested that there is a $10^7$ yr time delay between
the accumulation of gas by the M~51 spiral density wave and the formation of
self-gravitating clouds, resulting in the \thco\ emission being found downstream in the spiral arm where it is
also spatially correlated with H$\alpha$ emission.
\citet{meier04} find for NGC~6946 that the value of \r\ reaches high values of 40 away from the central
starburst which they attribute to diffuse, low-density molecular gas in and behind the molecular arms. 

Within the barred galaxy NGC~7479 spatial shifts between \thco\ and \twco\
have also been found \citep{huette00}. These shifts have been
attributed to density-wave or bar-like dynamical effects. The similarity
of these spatial shifts with those found in the Medusa may suggest
similar dynamical effects may be important. In particular, the transition
region between diffuse inter arm gas and self-gravitating
gas in the arms of M~51 suggesting that the similar shift in the Medusa is caused by the
funnelling and compression of gas towards the starburst region.

\subsection{Temperature gradient in the gas?}

In this scenario, the elevated value of \r\ in the eastern part of
the central region is caused not by diffuse molecular gas but by the
gas there being warmer and denser than in the north-west. For a temperature gradient to result in
elevated values of \r\ it is required that the average gas densities exceed
$3 \times 10^3$ $\cmmd$ (since the 1--0 transition of  \twco\ and \thco\
must be thermalised). A change of \r\ from 16 to $>$40 can be caused by a
temperature increase from 40 to 150 K in the gas (for a constant column-
and gas density per cloud). If both the temperature and gas density is
increasing towards the east then a smaller change in temperature could explain
the change in \r. If it is filled with warm and dense gas, the dust lane would be
more likely to harbour intense star formation
than the rest of the central region. However, as we mentioned above, the dust lane shows no sign 
of elevated star formation rates, although it is potentially hidden behind large masses
of dust. 
High resolution imaging of higher $J$-transition (e.g. 2--1) emission from
\thco\ and \twco\ will allow these diffuse gas and temperature effects to be
distinguished. If the faint \thco\ 1--0 emission in the eastern central part is caused by diffuse gas
the \thco\ 2--1 emission should be even fainter there while the opposite is true if
it is caused by a high gas temperature. This is because the 2--1 level would be more populated 
than the 1--0 for \thco\ if the gas is warm - while the opposite is true if the gas is diffuse.
One example where faint \thco\ 1--0 is accompanied by bright \thco\ 2--1 is the merger Arp~299
where it was suggested that the elevated value of \r\ is due to temperature effects \citep{aalto99}.
Other studies of multi-transition \thco\ can be found in \citet[e.g.][]{aalto95,glenn01,israel09}.
Additionally, high density tracer species such as HCN or HCO$^+$ will reveal where gas of densities $n > 10^4$ $\cmmd$
is located in relation to the starburst and dust lane. If the dust lane is filled with diffuse gas very
little dense gas will be found there. A high resolution dust SED will reveal where hot and cold
dust is in relation to the dust lane and central- and off-nuclear star formation.

\noindent
{\it An abundance gradient:} If the physical conditions in the gas are kept constant, an elevated \r\ may be
caused by a change in the $^{12}$C/$^{13}$C abundance ratio. This possibility has been discussed
by \citet[e.g.][]{casoli92} as an effect of low metallicity gas being transported from
the outskirts of the merger to its centre. The starburst would then contribute to
enriching the gas in $^{13}$C. In this scenario the gas in the dust lane has then not yet been
enriched by the starburst and there is therefore an east-west and north-south age-gradient. 
However, abundances in NGC~4194 are found to be close to solar, and no significant 
metallicity gradient can be found from a study of the properties of the star forming regions
\citep{hancock06}.

\subsection{The large scale gradient in \r}

In the Medusa, the lowest ratios of 7 -- 10 are found in the tail-like region north of the
centre. It is not unusual to find that \r\ is decreasing away from
the central region. This has been seen towards many starburst galaxies including
NGC~3256, NGC~1808 and large spiral galaxies such as M~51. 
The gas 10\arcsec\ north of the centre may therefore consist of rather ordinary cool Galactic-type
GMCs. This is interesting since the gas there does not appear to be in a normal disc
distribution, but rather in a shell or a tidal tail. There is no apparent star formation
going on in this region and the line widths are narrow.
If the gas in the central region is accumulating from the outer regions of the Medusa via the 
central dust lane, then our results suggest that the molecular cloud properties are 
normal in the outskirts, before becoming diffuse due to the funnelling of gas towards the centre and then accumulates
as self-gravitating, star-forming clouds in the centre of the Medusa.

\subsection{The nature of the Medusa starburst}

\subsubsection{Comparing with local minor mergers}

One of the few E+S minor mergers to have its molecular medium studied is the Medusa ``look-alike'' 
NGC~4441. Based on single dish observations of \twco\ and \thco\ 2--1 and 1--0 \citet{juette10}
suggest that the molecular gas is in an unbound, diffuse state. Interestingly NGC~4441 appears
to be an evolved version of the Medusa with similar morphology and luminosity and high resolution
\twco 1--0 observations of NGC~4441 show that it traces a centrally crossing dust lane. 
The star formation rate of NGC~4441 is much lower than that of the Medusa - likely due to the diffuse state of its molecular medium.
In this paper we suggest that a fraction of the Medusa molecular ISM may be the diffuse - it is possible that the {\it entire}
molecular gas content of NGC~4441 is in this state.
This raises interesting questions on the nature of the Medusa starburst and what it will evolve
into. Perhaps it will be left with a remnant, feeble dust lane, a small reservoir of diffuse gas
and only low level star formation, like NGC~4441.

\subsubsection{Is the star formation out of equilibrium?}

Despite being a minor merger of comparatively moderate luminosity, the Medusa
has an L$_{\rm FIR}$/L$_{\rm CO}$ ratio (163) similar to those of typical ULIRGs such as Arp~220 (210) 
- suggesting that its SFE rivals that of the compact ULIRGs.
However, the fraction of dense ($n > 10^4$ $\cmmd$) gas is significantly lower in the Medusa despite its similar
SFE to ULIRGs \citep{aalto00}. \citet{aalto00} estimate the gas consumption time for the Medusa
starburst to 40 Myrs and, if a significant fraction of the molecular medium is diffuse, this time could be even shorter.

The picture becomes even more interesting when one considers that most of the
ongoing star formation in the Medusa seems not to be traced by FIR or radio emission.
The H$\alpha$ SFR is found to be $\approx$46 M$_{\odot}$ yr$^{-1}$ \citep{hancock06}, while the FIR
estimated SFR is 6--7 M$_{\odot}$ yr$^{-1}$ \citep{aalto00}. (Note, that the SFR estimated
from H$\alpha$ would suggest an even more extreme SFE than the already high SFE derived from the FIR emission.)
It is possible that variations in the initial mass function (IMF) can account for some of these differences \citep[e.g.][]{wilkins08}, although the issue is far from resolved.  Such an anomaly in the 
form of a flatter upper IMF {\it could} lead to enhanced H$\alpha$ emission. 
The molecular morphology is also quite different from that of 1.4 GHz radio continuum - which is
also quite different from that found from other luminous galaxies where the two distributions 
are usually found to coincide. The optical star formation is concentrated in knots that could be precursors
to globular clusters. These knots have no clear spatial relation to either the molecular gas
distribution or the continuum. {\it This indicates a local deviation from the Kennicutt-Schmidt relation (KS) in
the Medusa that warrants further study. We suggest that the starburst of the Medusa is in an extreme transient phase
of very high efficiency.}

\subsubsection{Comparing the Medusa to high-redshifts extreme starbursts}

The Medusa merger has a number of features in common with some extreme starbursts 
at high-$z$ (despite its significantly lower luminosity). 
The starburst regions in high-$z$ systems often seem extended ($>$ 3 kpc), gas consumption times
are similarly short and there is a combination of ordered rotation and merger-driven random motions and inflows. 
More importantly, \citet{bothwell10} find for three $z$=3 ULIRGs a significant size difference between the CO distribution and
star formation tracers. They find that this size difference results in the SFEs within systems to vary by up to a factor of
five. As a consequence to their results they conclude that SMGs lie significantly above the KS relation, indicating that
stars may be formed more efficiently in these extreme systems than in other high-$z$ starburst galaxies. 
The Medusa merger is (by far) less luminous than the high-$z$ systems studied by \citet{bothwell10} but a
careful study of the extreme starburst of the Medusa merger may lead to insights that can be applied
to more distant galaxies.

\section{Conclusions}
We have obtained a high resolution map of the Medusa merger, NGC~4194,
in the \thco\ 1--0 line with the OVRO array. The main conclusions we draw from this map are:
   
   \begin{enumerate}
   
    \item The \thco\ 1--0 emission is mainly distributed in the north-western part of the 
    inner 8\arcsec\ region. There is also an off-centre \thco\ peak in the tail-like
    distribution to the north, where the CO/\thco\ line ratios are similar to those of galactic type GMCs
    
    \item The \twco/\thco\ 1--0 line ratio, \r, varies on a scale of 950 pc from 16 (typical values for starburst
	galaxies) in the north-west to $>$40 in the eastern part of the central region.
    
    \item We suggest that this shift between \twco\ and \thco\ emission is caused by the
	presence of diffuse gas in a prominent dust lane crossing the eastern part of the
	centre. A strong density wave-like phenomenon may cause shocks (hence the dust lane) and the \thco\ is tracing
    	more self-gravitating gas in the post-shock region in the starburst. In this scenario, the actual molecular mass 
	distribution is better traced by the \thco\ 1--0 emission than the \twco. Alternatively, the line ratio
    	shift is caused by temperature and density gradients in the gas. These two scenarios can be tested through imaging of
	\thco\ 2--1 and high density tracer molecules such as HCN and HCO$^+$.

	\item The extreme star formation efficiency and the lack of correlation between
	various star formation tracers suggest that the inner region of the Medusa merger is in a highly transient phase. 
	We predict that it will evolve into a quiescent object similar to NGC~4441, but that in
	its current evolutionary stage it may have some features in common with high redshift starburst galaxies.    

   \end{enumerate}

\begin{acknowledgements}
      This research has made use of the NASA/IPAC Extragalactic Database (NED) which is operated by the Jet Propulsion Laboratory, California Institute of Technology, under contract with the National Aeronautics and Space Administration.
\end{acknowledgements}

\bibliographystyle{aa}
\bibliography{13511}

\end{document}